\documentclass[]{spie}  %>>> use for US letter paper

\usepackage[]{graphicx}

\title{Phase referencing and narrow-angle astrometry in current and future interferometers} 

\author{Benjamin F. Lane\supit{a} and Matthew W. Muterspaugh\supit{a}
\skiplinehalf
\supit{a}MIT Center for Space Research, 70 Vassar St., Cambridge, MA. 02139, USA; 
}

\authorinfo{Further author information: (Send correspondence to B.F.L.)\\B.F.L.: E-mail: blane@mit.edu, Telephone: 1 617 253-3429 \\ M.W.M: E-mail: matthew1@mit.edu, Telephone: 1 617 547-1758
}
\begin{document} 
  \maketitle 

%%%%%%%%%%%%%%%%%%%%%%%%%%%%%%%%%%%%%%%%%%%%%%%%%%%%%%%%%%%%% 
\noindent
{\bf Copyright 2004 Society of Photo-Optical Instrumentation Engineers.\\}
This paper will be published in SPIE conference proceedings volume 5491, 
``New Frontiers in Stellar Interferometery.''  and is made available as 
and electronic preprint with permission of SPIE.  One print or electronic 
copy may be made for personal use only.  Systematic or multiple reproduction, 
distribution to multiple locations via electronic or other means, duplication 
of any material in this paper for a fee or for commercial purposes, or 
modification of the content of the paper are prohibited.

\begin{abstract}

Atmospheric turbulence is a serious problem for ground-based
interferometers. It places tight limits on both sensitivity and
measurement precision. Phase referencing is a method to overcome these
limitations via the use of a bright reference star.  The Palomar
Testbed Interferometer was designed to use phase referencing and so
can provide a pair of phase-stabilized starlight beams to a second
(science) beam combiner. We have used this capability for several
interesting studies, including very narrow angle astrometry. For close
(1-arcsecond) pairs of stars we are able to achieve a differential
astrometric precision in the range 20--30 micro-arcseconds.

\end{abstract}

%>>>> Include a list of keywords after the abstract 


%%%%%%%%%%%%%%%%%%%%%%%%%%%%%%%%%%%%%%%%%%%%%%%%%%%%%%%%%%%%%
\section{INTRODUCTION}
\label{sect:intro}  % \label{} allows reference to this section

Although they provide very high angular resolution, ground-based
stellar interferometers suffer from very limited sensitivity. This is
a consequence of atmospheric turbulence which corrupts the incoming
stellar wavefront on very short timescales and over small distances.
Typical numbers for atmospheric coherence length and time are $r_0 =
10 {\rm cm}$ and $\tau_0 = 2 {\rm ms}$ at visible wavelengths. An
interferometer must collect enough photons in a coherence volume
($\sim r_0^2\tau_0$) to detect fringes. Optimistically assuming a
system throughput of 10\%, this would place the S/N=3 tracking limit
at $\sim 8$th magnitude, assuming no detector noise. 

There are several ways to overcome this sensitivity limit. The most
obvious -- and most expensive -- is to put the interferometer in space
\cite{sim}, or at least at some location with favorable atmospheric
properties such as Antarctica\cite{domec}.  It is also crucial to maximize
system throughput and reduce other noise sources such as detector
read-noise\cite{l3ccd}.  It should also be noted that both $r_0$ and
$\tau_0$ scale approximately as $\lambda^{6/5}$ and thus going to
longer wavelengths will rapidly improve the situation. This is why the
current generation of interferometers often work in the near-IR.

In recent years, much work has been done to increase the apparent
$r_0$ via the use of adaptive optics\cite{ao} (AO); this has been
successful at both the Keck Interferometer\cite{keck} and the
VLTI\cite{vlti}. The use of AO-corrected apertures provides a
substantial improvement in collecting power, i.e. by a factor of $\sim
S \times (D/r_0)^2$, where $S$ is the AO Strehl ratio and $D$ is the
size of the AO-corrected aperture.  For a large aperture and a typical
AO Strehl one might expect an improvement by a factor of 100--300, as
long as a natural or laser guide star is available.

However, it is not clear that such an improvement will be sufficient
for certain areas of research, e.g. for imaging faint extra-galactic
objects or for narrow-angle astrometry\cite{shac92}. An additional
approach is to find a way to increase the effective atmospheric
coherence time, $\tau_0$. This can be done by simultaneously observing
a bright reference star and the science target using two separate
interferometric beam-combiners, and using the reference star to
determine the fringe motion introduced by the atmosphere. This is
known as phase referencing. By using a sufficiently fast control loop
($\sim 100$~Hz) it is possible to remove much of the fringe motion in
real time, and hence provide an artificially long coherence time for
the beam combiner used to observe the science target. In principle one
could consider extending the maximum allowed integration time from
milliseconds to seconds or even minutes, with a corresponding increase
in sensitivity. Of course, it should be noted that phase referencing,
like AO, requires a bright reference.

In this paper we discuss recent progress in the area of phase
referencing\cite{lac03} as applied to very narrow-angle differential
astrometry\cite{lm04}.  In particular, we have used the Palomar
Testbed Interferometer in a phase-referencing mode that allows us to
achieve very high precision relative astrometry between 0.1-arcsecond
pairs of stars; such astrometry would be useful in searching for
planets orbiting one of the components in the system. This
instrumental configuration can also be used to obtain simultaneous
high spectral and spatial resolution data through the use of
double-Fourier interferometry\cite{ridgway}.

The Palomar Testbed Interferometer\cite{pti} (PTI) is a long-baseline infrared
interferometer located at Palomar Observatory near San Diego,
California.  It typically operates in the H ($1.6 \mu{\rm m}$) and K
($2.2 \mu{\rm m}$) bands. It has three 40-cm diameter apertures
separated by 110-m and 80-m baselines, giving a maximum angular
resolution of $\sim3$ milli-arcseconds.

\section{PHASE REFERENCING}

Atmospheric turbulence above the interferometer can be characterized
by a distance over which the wavefront remains flat. This distance is
known the atmospheric coherence length. For typical observing sites,
we have
\begin{equation}
r_0 \simeq 0.1 \left(\frac{\lambda}{0.5 \mu{\rm m}}\right)^{6/5} {\rm m}
\end{equation}

In addition, since the atmosphere is not static, the wavefront errors
change on timescales of $\tau_0 \sim r_0/v$ where $v$ is a a wind
speed. The resulting phase perturbations obey an $f^{-8/3}$ power law.
An example of measured fringe motion is given in Figure 1.  The fringe
motion will tend to ``smear'' the fringe, reducing fringe visibility
and hence signal-to-noise ratio.

The phase fluctuations introduced by the atmosphere are correlated over
small angles, characterized by the angular isoplanatic angle $\theta_0$
such that
\begin{equation}
\left< \sigma_\theta^2 \right> = \left( \frac{\theta}{\theta_0} \right)^{5/3}
\end{equation}
where
\begin{equation}
\theta_0 = \left[ 2.914 k^2 (\sec \phi)^{8/3} \int C_n^2(z) z^{5/3} dz \right]^{-3/5}
\end{equation}
\noindent $C_n^2$ is a parameter representing the turbulence
amplitude, $k = 2\pi/\lambda$ is the wavenumber of the light and
$\phi$ is the zenith angle of the observation. 

 \begin{figure}
   \begin{center}
   \begin{tabular}{c}
   \includegraphics[height=7cm]{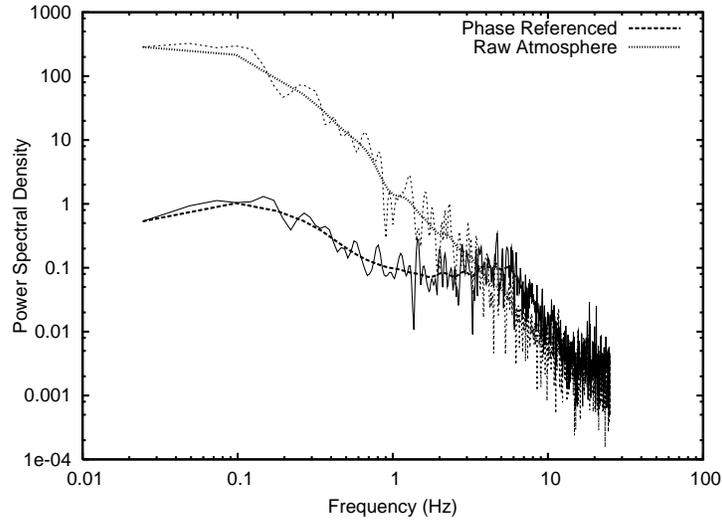}
   \end{tabular}
   \end{center}
   \caption[example] 
%>>>> use \label inside caption to get Fig. number with \ref{}
   { \label{fig:psd} Power spectral density of the measured fringe
phase, obtained using a fringe tracker at PTI, with and without phase
referencing. The atmosphere induces fringe motion that has a power-law
dependence ($P(f) \propto f^{-8/3}$)}
   \end{figure} 

If one has available an estimate of the fringe motion induced by the
atmosphere, it is sometimes possible to correct the phase and recover
the fringe signal. If the fringe is detected using an integrating detector,
or a detector with a non-zero read-noise, it is necessary to apply the
phase correction in real-time, i.e. by rapidly adjusting the
instrumental delay using a fast delay line or similar device.  The
phase reference can be obtained in many different ways; it can be
provided by a nearby reference star (located within the isoplanatic
patch, $\sim 30$ arcseconds in the near-IR), or it can be the science
target itself, observed at a different wavelength or with a different
beam-combiner (which may be optimized for phase measurement). It may
even be possible to provide a phase reference by measuring some
fraction of the atmospheric pathlength changes using a
laser\cite{townes}. In any case, the phase referencing system needs a
fast control loop to sense and correct the atmospheric pathlength
errors on millisecond timescales. The effect of atmospheric
turbulence, with and without active phase referencing is shown in Figure 2.

\begin{figure}
   \begin{center}
   \begin{tabular}{c}
   \includegraphics[height=7cm]{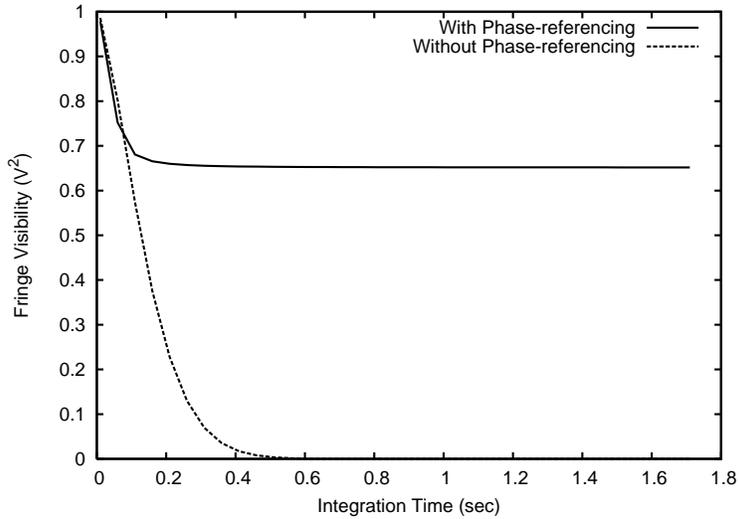}
   \end{tabular}
   \end{center}
   \caption[example] 
%>>>> use \label inside caption to get Fig. number with \ref{}
   { \label{fig:visloss} The effect of integration time on measured
fringe visibility, with and without phase referencing. Model 
parameters were appropriate for PTI, with a fringe tracker 
update rate of 100 Hz. On timescales shorter than the fringe-tracker
closed-loop bandwidth, phase referencing has little effect. 
However, it does remove the fringe motion at low frequencies, which 
would -- if left uncorrected -- completely wash out the fringe.}
   \end{figure} 

Clearly, phase referencing opens up a new regime for interferometry,
as beam-combination instruments can begin to work with exposure times
measured in seconds. This will allow the use of instruments that are
not necessarily optimized for short-exposure phase measurements but
rather high spectral resolution or high-precision visibility
measurements. This capability may also be applicable to multi-aperture
combination\cite{mozurk}, in that for an N-aperture interferometer it
is only necessary to fringe-track on N baselines in order to
keep the array ``phased''. The full set of N(N-1)/2 baselines
can then be combined in a different, dedicated beam-combiner.

\section{ASTROMETRY}

One application that can take advantage of phase referencing is
narrow-angle astrometry. In this case one is interested in measuring
the angle between a close pair of stars to very high precision; such
astrometry can be used both to determine orbital parameters (which can
be combined with radial velocity and photometry data to determine
component masses, distances and luminosities) to high precision, and to
search for planets in the system.

PTI is unique among current interferometers in that it has a
``dual-star'' configuration, which makes it possible to simultaneously
track two close (separation $\le 60$ arcseconds) stars. To achieve
this, PTI is equipped with two separate optical trains and
beam-combiners that share the same apertures.  PTI is also
particularly suited to phase-referencing in that it is capable of
fringe-tracking at bandwidths high enough to follow atmospheric
motion, and hence one beam combiner can track and correct the
atmosphere while the other can be used for high-precision
measurements.

For close pairs of stars, astrometric errors introduced by the 
atmosphere will tend to be common-mode, and hence will subtract 
out of a differential measurement. Shao \& Colavita\cite{shac92}
quantified this effect (in arcseconds of astrometric error) as
\begin{equation}
\label{eqn:astro}
\sigma_{a} = 540 {\rm B}^{-2/3} \theta t^{-1/2}
\end{equation}
\noindent where B is the baseline in meters, $\theta$ is the angular
separation (in radians) of the stars, and $t$ the integration time in
seconds.

\begin{figure}[!ht]
   \begin{center}
   \begin{tabular}{c}
   \includegraphics[height=7cm]{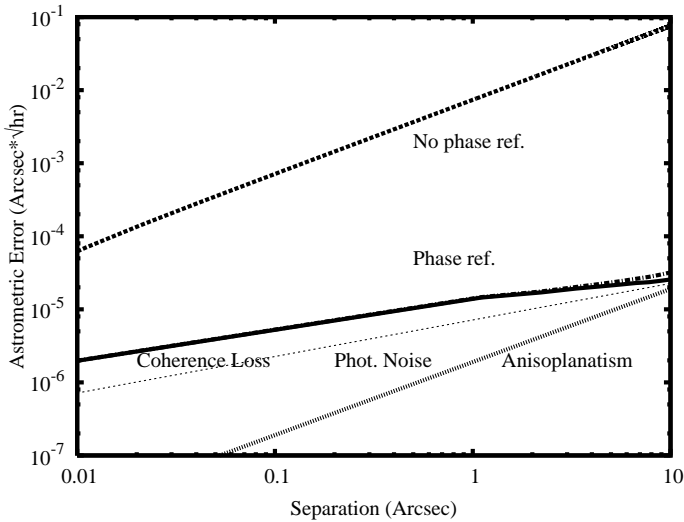}
   \end{tabular}
   \end{center}
   \caption[example] 
%>>>> use \label inside caption to get Fig. number with \ref{}
   { \label{fig:naa} The
expected narrow-angle astrometric performance in arcseconds for the
phase-referenced fringe-scanning approach, for a fixed scan
rate. There are three primary sources of astrometric error in this
method: angular anisoplanatism\cite{shac92}, temporal
decoherence\cite{lac03}, and photon noise. Also shown is
the magnitude of the temporal decoherence effect in the absence of
phase referencing, illustrating why stabilizing the fringe via phase
referencing is necessary. }
\end{figure} 

From Equation \ref{eqn:astro} it is clear that at small separations
the atmospheric error becomes negligible. This motivated us to modify
PTI so as to allow us to study binaries with separations on the order
of 0.1--1 arcseconds, i.e.  larger than the nominal interferometric
field of view (given by $\lambda/\Delta\lambda \times \lambda/B$), but
smaller than the 30--60 arcseconds previously considered for
narrow-angle astrometry\cite{shac92}. The probability of finding a
background reference star in this reduced field is very small, but
outweighed by the fact that binary stellar systems with these apparent
separations are quite common.

For a binary separation of 0.1 arcsecond and an interferometric
baseline of 100 meters, the predicted error due to the atmosphere is
$\sim 0.2 \mu$arcsecond in an hour. Of course, there are additional sources of
astrometric error, including photon noise and potential systematic
errors associated with the instrument. In particular, if the measurements
of the two stars are not truly simultaneous -- as in this case, see below -- 
there will be a small error associated with fringe motion during the 
time between the measurements. Nevertheless, we estimate that
PTI should be able to obtain an astrometric precision at the level of
$10 \mu$arcseconds in an hour of observation (Figure \ref{fig:naa}).

\subsection{Instrument Configuration}
 
\begin{figure}[t]
   \begin{center}
   \begin{tabular}{c}
   \includegraphics[height=7cm]{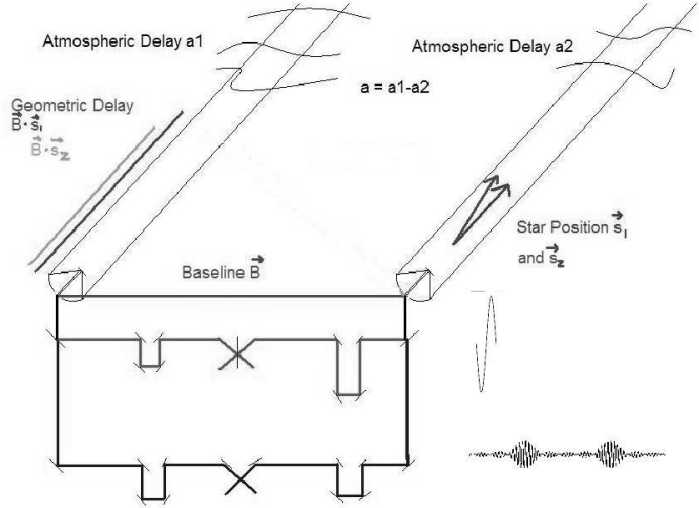}
   \end{tabular}
   \end{center}
   \caption[example] 
%>>>> use \label inside caption to get Fig. number with \ref{}
   { \label{fig:config} The instrument configuration for
phase-referenced narrow-angle astrometry. One beam-combiner is used to
track and correct for atmospheric turbulence; corrections it derives 
are applied to both delay lines. The second beam combiner 
scans over $\sim 200 \mu$m in differential optical path, tracing out
the double-fringe packet produced by a close binary star. }
\end{figure} 

We configured PTI to observe close (separation less than 1 arcsecond)
visual binaries, so-called ``speckle binaries''. In this configuration
(Figure \ref{fig:config}), the incoming starlight was split 70/30
between a fringe tracker running at 100 Hz and a second, astrometric
beam combiner. The astrometric beam combiner did not track the fringe
but rather it measured the intensity of the combined stellar beams
while the relative optical path between the arms of the interferometer
was modulated (using a differential delay line) in a triangle-wave
pattern with an amplitude of $\sim 100 \mu$m and a period of $\sim
1$Hz. An example of this data is given in Figure \ref{fig:waterfall}.

\subsection{Data Processing}

\begin{figure}[h]
   \begin{center}
   \begin{tabular}{c}
   \includegraphics[height=7cm]{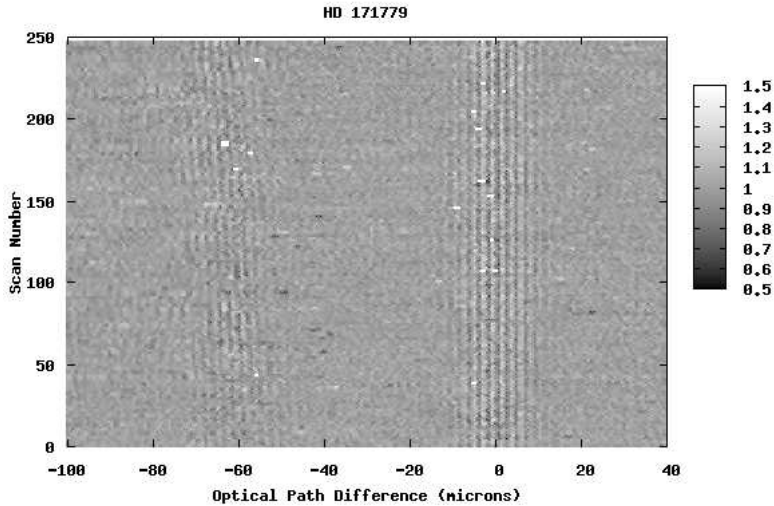}
   \end{tabular}
   \end{center}
   \caption[example] 
%>>>> use \label inside caption to get Fig. number with \ref{}
   { \label{fig:waterfall} Measured intensity in the detector as a
function of differential optical path, for successive scans of the
speckle binary system HD 171779. Each scan takes 1.5 seconds to
acquire. The fringe tracker was locked on to the bright star (around
0), while the second star produces a fringe pattern which starts at -60
$\mu$m and moves due to baseline rotation. Although the second
fringe pattern is relatively faint, the effect of coherently 
co-adding 500--2000 scans produces a high signal-to-noise 
ratio in the final astrometric measurement.}
\end{figure} 

We extract the relative astrometric positions from data such as that
shown in Figure \ref{fig:waterfall} as follows. First, we construct a
search grid in differential R.A. and declination over which to search.
For each point in the search grid we calculate the expected
differential delay based on the interferometer location, baseline
geometry, and time of observation for each scan. A model of a
double-fringe packet is then calculated and compared to the observed
scan to derive a $\chi^2$ value; this is repeated for each scan,
co-adding all of the $\chi^2$ values associated with that point in the
search grid. The final $\chi^2$ surface as a function of differential
R.A. and declination is thus derived. The best-fit astrometric
position is found at the minimum-$\chi^2$ position, with uncertainties
defined by the appropriate $\chi^2$ contour -- which depends on the
number of degrees of freedom in the problem and the value of the
$\chi^2$-minimum. Because the data was obtained with a single-baseline
instrument, the resulting error contours are very elliptical, with
aspect ratios at times $\ge 10$.

One potential complication with fitting a fringe to the data is that
there are many local minima spaced at multiples of the operating
wavelength. If one were to fit a fringe model to each scan separately
and average (or fit an astrometric model to) the resulting delays, one
would be severely limited by this fringe ambiguity (for a 110-m
baseline interferometer operating at $2.2 \mu$m, the resulting
positional ambiguity is $\sim 4.6$ milli-arcseconds). However, by
using the $\chi^2$-surface approach, and co-adding the probabilities
associated with all possible delays for each scan, the ambiguity
disappears. This is due to two things, the first being that co-adding
simply improves the signal-to-noise ratio. Second, since the
observations usually last for an hour or even longer, the associated
baseline change due to Earth rotation also has the effect of ``smearing''
out all but the true global minimum. The final $\chi^2$-surface does
have dips separated by $\sim 4.6$ milli-arcseconds from the true 
location, but they only show up at the 40-$\sigma$ contour level.

\subsection{Preliminary Results}
\begin{figure}[h]
   \begin{center}
   \begin{tabular}{c}
   \includegraphics[height=10cm]{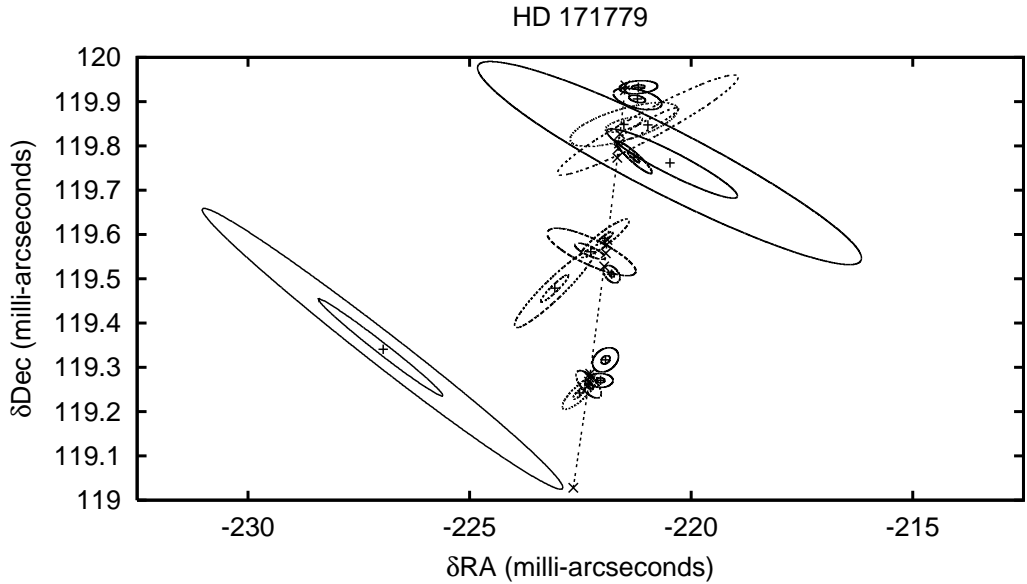}
   \end{tabular}
   \end{center}
   \caption[example] 
%>>>> use \label inside caption to get Fig. number with \ref{}
   { \label{fig:astrores} Measured differential astrometry between the
components of the speckle binary HD 171779, together with a best-fit
orbital motion curve. The contours represent the formal 1-$\sigma$ and
4-$\sigma$ uncertainties. Note that the single-baseline configuration
used to obtain this data is sensitive primarily in the declination
direction, and therefore the Declination-axis has been magnified by a
factor of 10 compared to the R.A. axis. This data implies a
night-to-night astrometric repeatability of 24 $\mu$-arcseconds in the
direction of greatest sensitivity (approximately aligned to the
declination axis), and a somewhat less impressive 470 $\mu$-arcsecond
repeatability in the orthogonal direction. }
   \end{figure} 

We have obtained exploratory data on 25 binaries, and have intensively
followed the binary system HD 171779 in order to evaluate the
performance and repeatability of our system (Figure
\ref{fig:astrores}).  HD 171779 is a pair of K-giants in a long-period
orbit, such that the apparent motion is nearly linear over the few
months of observation. A typical observation of this pair takes an
hour, and produces a formal 1-$\sigma$ error ellipse approximately $5
\times 100$~ $\mu$-arcseconds in size; the smaller axis being aligned
with the mean baseline orientation (which in turn depends on the exact
timing of the observation). A linear two-dimensional (taking into
account the full shape of the error ellipse) fit of 15 nights of data
produces a reduced $\sqrt{\chi_r^2}= 4.7$, indicating that the formal
errors underestimate the true errors by a factor of 4.7. We note that
fits to 1-D projections (e.g. along the declination or R.A. axis) show
similar values of reduced $\chi^2$, indicating that the error scaling
is uniform.  Hence we conclude that the night-to-night repeatability
is 24 and 470 $\mu$-arcseconds in the minor and major axes
respectively. We note that we have recently improved aspects of the
instrument metrology system and data processing, and expect to achieve
the 10 $\mu$-arcsecond goal in the next few months.

\section{DISCUSSION}

\begin{figure}[h]
   \begin{center}
   \begin{tabular}{c}
   \includegraphics[height=7cm]{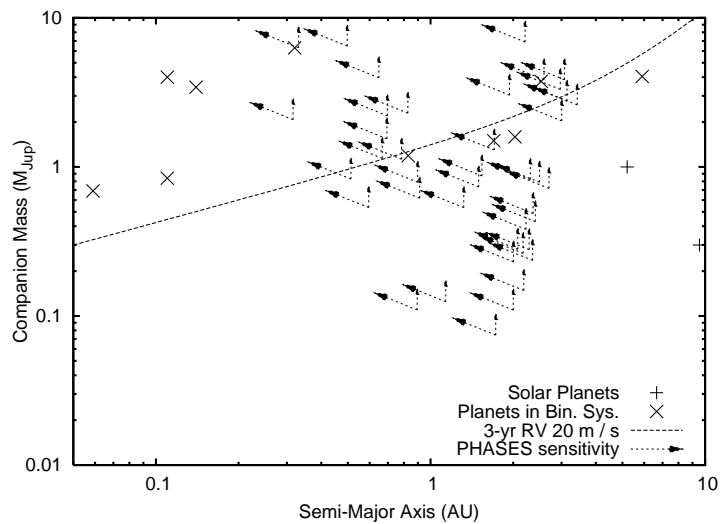}
   \end{tabular}
   \end{center}
   \caption[example] 
%>>>> use \label inside caption to get Fig. number with \ref{}
   { \label{fig:planets} 
Phase space explored by radial velocity and
astrometry. The figure shows the areas in the semi-major axis
vs. companion mass plane explored by a notional astrometry project
(``PHASES''), for each specific target in our sample; astrometry will
find any planet located above and to the left of the arrows. This
accounts for the actual mass and distance to the binaries in question.
Also shown are a number of planets known to exist in wide (2 arcsec
separation) binary systems.  The precision for the RV sample is 20 m
s$^{-1}$ which is a realistic value given the additional photon noise
and systematics present for compact binaries.  For binary separations
of 10 AU, companions in orbits inside approximately 1 AU are expected
to be stable; the stability zone grows with increasing binary
separation.

}
   \end{figure}  

Ground-based interferometers are severely challenged by the effects of
atmospheric turbulence. The short integration times required in the
absence of any active compensation limits interferometry to all but the
brightest astronomical sources, and limits the types of observations
that can be done (spectral resolution, multi-aperture combination
etc). One solution is to divide the problem into two parts which can
be optimized separately. A phase-tracking beam-combiner can use either
a nearby reference source or the science target itself -- if it is
sufficiently bright -- to stabilize the phase of the incoming stellar
wavefront. A second, science-optimized beam combiner can then obtain
desired data at a relatively leisurely pace. 

We have used a phase-referencing approach to observe close pairs of
binaries (separations less than 1 arcsecond) in order to obtain very
high precision differential astrometry. Over 15 nights of data on the
binary HD 171779 we find a night-to-night repeatability of
approximately 24 $\mu$-arcseconds in one axis. This level of
astrometric precision is sufficient to begin a search for Jupiter-mass
planets around the components of nearby binary stellar systems, given
that simulations\cite{hw} indicate there are regions where planets can
persist indefinitely (i.e. sufficiently close to one of the stellar
components). A dedicated astrometric survey could quickly answer the
question of whether or not planets form in these comparatively close
systems. We have begun such a survey, PHASES (``Palomar High-precision
AStrometric Exoplanet Search'') and hope to have results in the next
few years.

%%%%%%%%%%%%%%%%%%%%%%%%%%%%%%%%%%%%%%%%%%%%%%%%%%%%%%%%%%%%%
\acknowledgments %>>>> equivalent to \section*{ACKNOWLEDGMENTS}
 
It is a pleasure to thank M. Colavita, S. R. Kulkarni, B. F. Burke,
M. Konacki, A. Boden, N. Safizadeh and K. Rykoski for their
contributions to the astrometry effort. Observations with PTI are made
possible through the efforts of the PTI Collaboration, which we
acknowledge. Interferometer data was obtained at the Palomar
Observatory using the NASA Palomar Testbed Interferometer, supported
by NASA contracts to the Jet Propulsion Laboratory. This research has
made use of the Simbad database, operated at CDS, Strasbourg,
France. MWM acknowledges the support of the Michelson Graduate
Fellowship program. BFL gratefully acknowledges support from a
Pappalardo Fellowship in Physics.

%%%%%%%%%%%%%%%%%%%%%%%%%%%%%%%%%%%%%%%%%%%%%%%%%%%%%%%%%%%%%
%%%%% References %%%%%

\bibliography{report}   %>>>> bibliography data in report.bib
\bibliographystyle{spiebib}   %>>>> makes bibtex use spiebib.bst

\end{document}